\documentclass[10pt,twocolumn]{IEEEtran}
\usepackage{xspace,amsthm,amsmath,amssymb,amsfonts,epsfig,syntonly}
\usepackage{cite,bm,color,url,textcomp,balance}
\usepackage{epstopdf}
\usepackage{empheq}

\usepackage[linesnumbered,ruled,vlined]{algorithm2e}
\usepackage{algpseudocode}
\allowdisplaybreaks[3]

\usepackage{graphicx}{\large}
\usepackage{subfigure}
\usepackage{booktabs, multicol, multirow, makecell}

\usepackage{amsmath,amssymb,amsthm}
\usepackage{bigints}
\usepackage{stfloats,siunitx}

\usepackage{comment}
\usepackage{textcomp}

\definecolor{mygreen}{RGB}{32,178,170}  

\long\def\symbolfootnote[#1]#2{\begingroup
\def\thefootnote{\fnsymbol{footnote}}
\footnote[#1]{#2}\endgroup}

\psfull

\allowdisplaybreaks[4]

\IEEEoverridecommandlockouts

\begin{document}
\title{
RIS-Assisted Jamming Rejection and Path Planning for UAV-Borne IoT Platform: A New Deep Reinforcement Learning Framework
}

\author{Shuyan Hu,~\IEEEmembership{Member, IEEE}, Xin Yuan,~\IEEEmembership{Member, IEEE},
Wei Ni,~\IEEEmembership{Senior Member, IEEE}, \\ Xin Wang,~\IEEEmembership{Fellow, IEEE}, and Abbas Jamalipour,~\IEEEmembership{Fellow, IEEE}
\thanks{Shuyan Hu and Xin Yuan contributed equally to this work.\par
S. Hu and X. Wang are with the Department of Communication Science and Engineering, Fudan University, Shanghai 200433, China (e-mails: \{syhu14, xwang11\}@fudan.edu.cn).\par
X. Yuan and W. Ni are with CSIRO Data61, Sydney, NSW 2122, Australia (e-mails: \{xin.yuan, wei.ni\}@data61.csiro.au).\par
A. Jamalipour is with the School of Electrical and Information Engineering, The University of Sydney, Sydney,
NSW 2006, Australia (email: a.jamalipour@ieee.org).\par
Corresponding author: X. Wang.\par

}}

\maketitle

\begin{abstract}

This paper presents a new deep reinforcement learning (DRL)-based approach to the trajectory planning and jamming rejection of an unmanned aerial vehicle (UAV) for the Internet-of-Things (IoT) applications. Jamming can prevent timely delivery of sensing data and reception of operation instructions. With the assistance of a reconfigurable intelligent surface (RIS), we propose to augment the radio environment, suppress jamming signals, and enhance the desired signals. The UAV is designed to learn its trajectory and the RIS configuration based solely on changes in its received data rate, using the latest deep deterministic policy gradient (DDPG) and twin delayed DDPG (TD3) models. Simulations show that the proposed DRL algorithms give the UAV with strong resistance against jamming and that the TD3 algorithm exhibits faster and smoother convergence than the DDPG algorithm, and suits better for larger RISs. This DRL-based approach eliminates the need for knowledge of the channels involving the RIS and jammer, thereby offering significant practical value.

\end{abstract}

\section{Introduction}

The use of unmanned aerial vehicles (UAVs) has been growing in popularity with the expansion of the Internet of Things (IoT)~\cite{iotjdrone,iotjthrough,Li2016Energy}.
For instance, UAVs equipped with sensors and cameras are increasingly deployed to monitor and collect data on air quality, traffic, and other environmental factors for IoT applications~\cite{iotjtask,Li2019On}. The data collected by the UAVs need to be transmitted back to a control center or ground base station (BS) for data analysis or for triggering automated responses, such as adjusting traffic lights to alleviate congestion~\cite{hu21}.

Jamming can have severe effects on the physical-layer security of UAV-borne IoT applications~\cite{yuan2019Secrecy, hu20,iotjdang,Li2019Energy}. An attacker can use jamming equipment to disrupt the wireless transmissions between the UAVs and the BS, preventing the UAVs from transmitting their sensing data and receiving operation instructions in a timely manner and causing the UAVs to fail their missions~\cite{iotjsafeguard}. As a matter of fact, jamming has been identified to be one of the most critical threats to the massive IoT connections in the context of the upcoming sixth-generation (6G) communication systems~\cite{iotjnoma}.

Identifying and eliminating jammers is challenging without specialized equipment, such as radar~\cite{yuan20}. Previous research on jamming cancellation for UAV-borne IoT applications has not specifically focused on UAV-borne IoT platforms. In~\cite{duo20}, the authors targeted to improve the worst-case long-term data rate for a UAV-assisted wireless sensor network under jamming attacks by jointly configuring the UAV's transmission strategy and 3D flight path. In~\cite{access}, a reinforcement learning (RL)-based communication strategy was developed between a UAV swarm and a BS to counteract jamming attempts. This improved the communication quality of UAVs by exploring the motion and antenna spatial domain. In~\cite{lizhiwei21}, the authors proposed a secure UAV communication scheme against smart jammers using a knowledge-based RL approach, which leveraged domain information to reduce the state space and speed up the convergence of the RL algorithm.

On the other hand, reconfigurable intelligent surfaces (RISs) were developed as a means of creating programmable wireless transmission environments~\cite{huang20}. These surfaces, made up of passive reflecting units with reprogrammable phase shifts~\cite{cao21}, can be placed on building surfaces, and are expected to be an integral part and effective enhancer for the IoT. By configuring the phase shifts, the RIS-reflected signals are added constructively to the direct signal to improve signal quality or destructively to reduce interference~\cite{sunchao21}. However, the potential of RISs for anti-jamming applications has not been widely studied~\cite{qq21, liuyw21}. In~\cite{qq19twc}, a collective active and passive beamforming design was formulated to lower transmit power by optimizing the continuous phase shifts of an RIS. The result was later extended to discrete phase shifts in~\cite{qq20phase}. 
The RIS-assisted secure transmission was studied in~\cite{iotjamm, iotjsecure}, where fast RL and deep RL (DRL) were used to design the active and passive beamformers. For instance, in~\cite{iotjamm}, jamming signals were modulated and reflected by an RIS to reduce the eavesdropping data rate.

As far as we know, the RIS-assisted jamming rejection has yet to be addressed in the context of UAV-borne IoT platforms. While there have been studies that have examined general RIS-assisted UAV communications, e.g., in~\cite{iotjcomp,li2020, Wang2015VANET, iotjurllc, liux21}, none have considered the impact of jamming attacks. For example, in~\cite{li2020}, the authors jointly optimized UAV flight path and RIS passive beamformers to achieve the largest average rate of the terrestrial user. In~\cite{iotjurllc}, a UAV and RIS were configured to deliver ultra-reliable and low-latency commands among terrestrial IoT devices using nonconvex optimization. These studies are inapplicable to the RIS-assisted jamming rejection for UAV-borne IoT platforms.

In this paper, we put forth a new DRL-based architecture for the flight path planning and jamming cancellation of UAV-borne IoT platforms. A fixed-wing UAV is considered. Our architecture utilizes an RIS, which dynamically modifies the wireless transmission environment to mitigate jamming power and enhance intended signals to the UAV. To accomplish this, we devise a new Deep Deterministic Policy Gradient (DDPG) model and its enhancement, Twin-Delayed DDPG (TD3), to allow the UAV to learn its flight path and the RIS configuration based solely on its received data rate, eliminating the need for the channel state information (CSI) involving the RIS and jammer in the flight path training. This presents a significant practical advantage, as the estimation of CSI involving an RIS is complex and may not be able to be performed in real-time.

The main contributions of this paper are as follows:
\begin{itemize}
\item[$\bullet$] A new problem is introduced to jointly optimize flight path planning and RIS-assisted jamming cancellation  to maximize the data rate of a UAV-borne IoT platform. 
The problem is non-straightforward for its non-convexity and sequential decision-making nature.

\item[$\bullet$] A new DRL architecture is proposed to solve the new problem and allow the UAV to learn its flight path and the RIS configuration based solely on its received data rate, eliminating the need for CSI in the training process.

\item[$\bullet$] The DRL architecture is implemented using the latest DRL models, DDPG and its twin-delayed version, i.e., TD3. 
While TD3 is generally applicable to the problem under investigation, DDPG can benefit from its simpler network architecture and smooth convergence and is suitable for problems with smaller scales, e.g., fewer RISs, and less stringent mission time requirements.

\end{itemize}
%
The proposed DRL approach is validated through extensive simulations, showing its exceptional resistance against jamming. 
Particularly, the DDPG demonstrates faster and smoother convergence when the mission time is long and the RIS is small. The TD3 outperforms the DDPG in robustness against the jammer's position, especially when the mission time is short. This is crucial, as accurately locating the jammer and estimating its CSI can be practically challenging.

The remainder of this paper is arranged as follows. In Section~\ref{sec.model}, the system model is described. In Section~\ref{sec.prob}, we formulate the problem of jointly designing the UAV's flight path and the RIS configuration to maximize the data rate in the presence of an unknown jammer. In Section~\ref{sec.ddpg}, we propose new DRL solutions to the problem. Performances are gauged in Section~\ref{sec.sim}. The paper is concluded in Section~VI.

\emph{Notation}: Boldface lower- and upper-cases indicate vectors and matrices, respectively;
$\mathbb C^N$ denotes the space of $N \times 1$ complex-valued column-vectors;
$\| \cdot \|$ denotes the Euclidean norm;
$(\cdot)^T$ and $(\cdot)^H$ stand for transpose and conjugate transpose, respectively;
diag$(\mathbf a)$ is a diagonal matrix with the elements of $\mathbf a$ along the diagonal;
$\otimes$ stands for the Kronecker product.

\begin{figure}[t]
\centering
\includegraphics[width=0.47\textwidth]{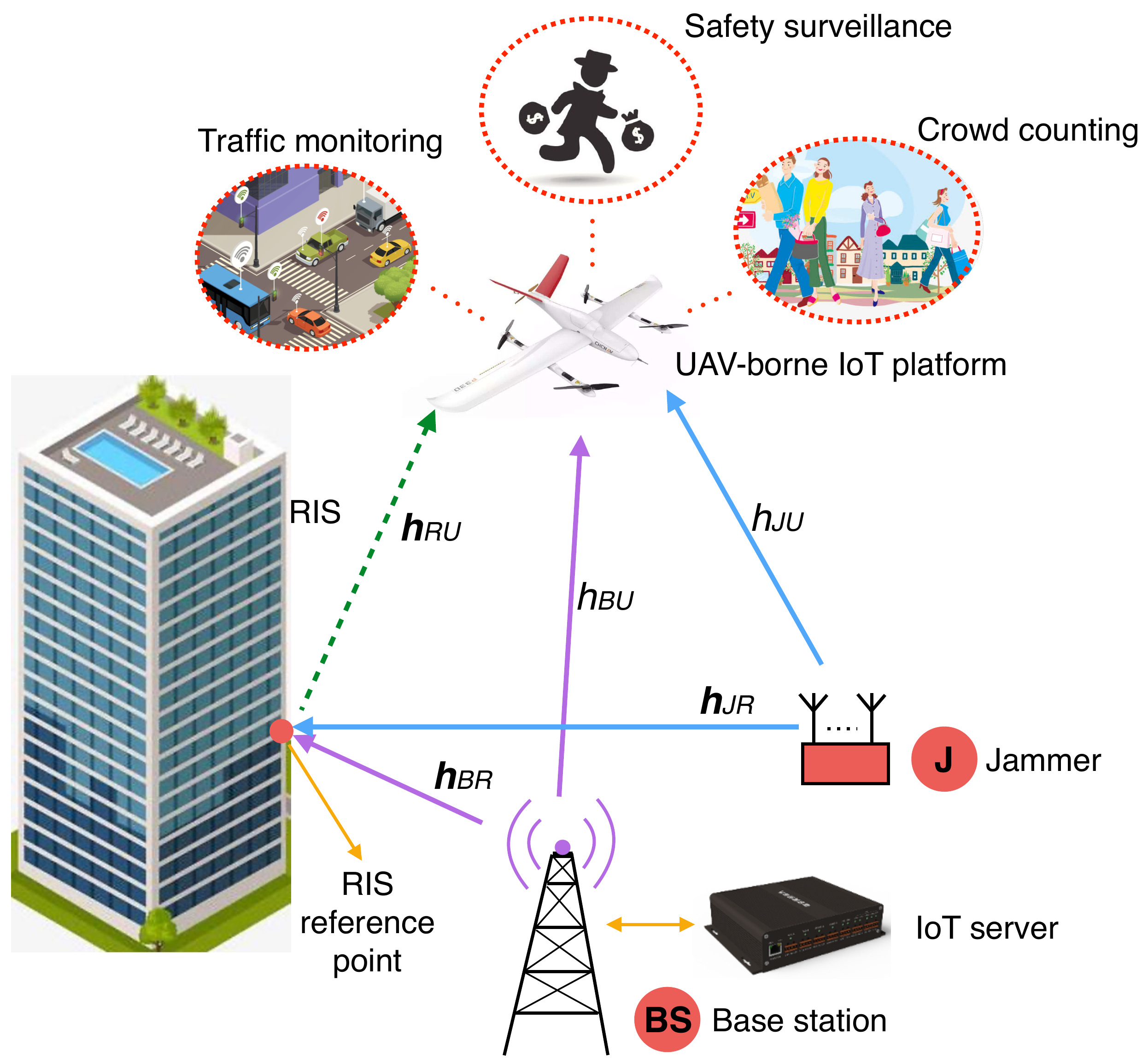}
\caption{An RIS-assisted anti-jamming UAV-borne IoT platform, where an RIS is adaptively configured along with the UAV trajectory to enhance the desired signals and reject the jamming signals.}
\label{systmod}
\end{figure}

\section{System Model}\label{sec.model}

As depicted in Fig. \ref{systmod}, a ground BS offers wireless communication service for a fixed-wing UAV in the presence of a terrestrial jammer.
Consider a three-dimensional (3D) Cartesian coordinate system.
The BS and the jammer are placed at $\boldsymbol q_{\rm B}=[0, 0, 0]^T$ and
$\boldsymbol q_{\rm J}=[x_{\rm J}, y_{\rm J}, 0]^T$.
An RIS is deployed to facilitate the UAV communication and suppress the jamming signals from the jammer.
We suppose that the BS, UAV and jammer all have a single antenna for description convenience
(but the proposed DRL architecture can be extended to support multi-antenna BSs and UAVs, as part of our future work).
Only based on its received data rate, the UAV determines its flight path and the RIS configuration dynamically.
The RIS is installed with a smart controller to procure the command from the UAV (via the cellular system)
for RIS configuration~\cite{muxd21}.

\subsection{UAV Mobility Model}\label{sec.uavmob}

The UAV functions for a finite scheduling horizon of $T$ seconds, which is split
into $T_w$ time slots indexed by $t$, and $t=1, \cdots, T_w$.
A slot lasts $\delta=T/ T_w$, which is short enough that the UAV can be viewed as stationary per slot.
The UAV is traveling from a predetermined starting position $\boldsymbol q_0=[x_0, y_0, z_0]^T$ to a predetermined ending position
$\boldsymbol q_F=[x_{F}, y_{F}, z_F]^T$. 
The UAV's 3D coordinates are $\boldsymbol q_t=[x_t,y_t, z_t]^T, \forall t$.
Let $\boldsymbol V_t:=[V_{xt}, V_{yt}, V_{zt}]^T$ and $\boldsymbol A_t:=[A_{xt}, A_{yt}, A_{zt}]^T$ collect the velocity and acceleration
of the UAV per slot $t$, respectively.
The UAV obeys some mobility constraints~\cite{sun21}:
\begin{subequations}\label{eq.mobfix}
\begin{align}
&\boldsymbol q_{t+1}=\boldsymbol q_t + \boldsymbol V_t \delta +\frac{1}{2} \boldsymbol A_t \delta^2,~\forall t, \label{eq.mobfix1}\\
&\boldsymbol q_{F}=\boldsymbol q_{T_w}, \label{eq.mobfix2}\\
&\boldsymbol V_{t+1}=\boldsymbol V_t + \boldsymbol A_t \delta, ~\forall t, \label{eq.mobfix3}\\
& |A_{i,t}| \le A_{\max}, i \in \{x,y,z\}, ~\forall t,  \label{eq.mobfix4}\\
&V_{\min} \le \| \boldsymbol V_t \|  \le V_{\max}, ~\forall t, \label{eq.mobfix5} \\
& V_{zt}/ \| \boldsymbol V_t \|  \le \sin \vartheta, ~\forall t, \label{eq.mobfix6}
\end{align}
\end{subequations}
where $A_{\max}$ is the UAV's largest acceleration;
$V_{\max}$ and $V_{\min}$ are the UAV's maximum and minimum speeds;
and $\vartheta$ is the largest UAV pitch angle when ascending or descending.

\subsection{RIS Configuration}\label{sec.irs}
The RIS is adhered on the outer surface of a building, which is on the $(x,z)$-plane and aligns with the $x$-axis. 
The RIS has a uniform rectangular array (URA) of $N=N_x N_y$ reflecting elements (units),
and a controller to dynamically control the phase shift of each unit.
Let $\boldsymbol \Theta_t := \text{diag} (e^{j\theta_1^t}, \ldots, e^{j\theta_N^t})$ be the phase shift matrix
for the RIS at time slot $t$,
where $\theta_n^t = \theta_{(n_x-1)N_y + n_y}^t \in [-\pi, \pi), n=1, \ldots, N$,
is the phase shift of the $n$-th reflecting unit which is located at the $n_y$-th row and the $n_x$-th column of the RIS, and
$j=\sqrt{-1}$.
The first element of the RIS is at the right bottom corner of the RIS,
and its coordinates are $\boldsymbol q_{\rm R}=[x_{\rm R}, y_{\rm R}, z_{\rm R}]^T$.

\subsection{Channel Model}\label{sec.channel}

The distances of the BS-UAV link $d_{\rm BU}^t$, the BS-RIS link $d_{\rm BR}$, 
the Jammer-RIS link $d_{\rm JR}$, the Jammer-UAV link $d_{\rm JU}^t$, and RIS-UAV link $d_{\rm RU}^t$ are given by
\begin{subequations}
\begin{align}
d_{\rm BU}^t&=\| \boldsymbol q_t \|, ~\forall t, ~d_{\rm BR}=\| \boldsymbol q_{\rm R} \|,~d_{\rm JR}=\| \boldsymbol q_{\rm J} - 
\boldsymbol q_{\rm R} \|, \\
d_{\rm JU}^t&=\| \boldsymbol q_t-\boldsymbol q_{\rm J}  \|, ~ d_{\rm RU}^t=\| \boldsymbol q_t-\boldsymbol q_{\rm R}  \|, ~\forall t.
\end{align}
\end{subequations}

Consider an LoS channel for the RIS-UAV link (i.e., the R-U link),
and Rician fading channels between the BS/jammer and the UAV (i.e., the B-U and J-U links),
and between the BS/jammer and the RIS (i.e., the B-R and J-R links).
The channel gains of the B-U and J-U links are: $\forall t$,
\begin{subequations}
\begin{align}
h_{\rm BU}^t&=\sqrt{\rho (d_{\rm BU}^t)^{-\kappa_1}} \left(\sqrt{\frac{\beta_t}{1+\beta_t}} g_{\rm BU}^t 
+ \sqrt{\frac{1}{1+\beta_t}} \tilde g_{\rm BU}^t \right),  \\
h_{\rm JU}^t&=\sqrt{\rho (d_{\rm JU}^t)^{-\kappa_1}} \left(\sqrt{\frac{\beta_t}{1+\beta_t}} g_{\rm JU}^t 
+ \sqrt{\frac{1}{1+\beta_t}} \tilde g_{\rm JU}^t \right), 
\end{align}
\end{subequations}
where $\rho$ stands for the path loss at the reference distance $d_0=1$~m with the path loss exponent $\kappa_1>2$;
$\beta_t$ is the Rician factor of the B-U and J-U links;
$g_{\rm BU}^t$ and $g_{\rm JU}^t$ are the deterministic LoS components with $| g_{\rm BU}^t | =1$ and $| g_{\rm JU}^t | =1$;
$\tilde g_{\rm BU}^t$ and $\tilde g_{\rm JU}^t$ stand for stochastic dispersion captured by a zero-mean, unit-variance
circularly symmetric complex Gaussian (CSCG) random variable.
The elevation-angle-reliant Rician factor $\beta_t$ is captured by the following exponential function~\cite{you19}
\begin{equation}
\beta_t = \xi_1 \exp \left(\xi_2 \arcsin(z_t/d_{\rm BU}^t) \right), ~\forall t,
\end{equation}
where $\xi_1$ and $\xi_2$ are two constant coefficients dependent on the environment.

The channel gains of the B-R, J-R, and R-U links, denoted by $\boldsymbol h_{\rm BR}\in {\mathbb C}^{N \times 1}$,
$\boldsymbol h_{\rm JR}\in {\mathbb C}^{N \times 1}$, and $\boldsymbol h_{\rm RU}^t \in {\mathbb C}^{N \times 1}$, are
\begin{subequations}
\begin{align}
\boldsymbol h_{\rm BR}&=\underbrace{\sqrt{\rho d_{\rm BR}^{-\kappa_2}}}_{\text{path loss}}
\underbrace{\Bigg( \sqrt{\frac{\beta}{1+\beta}} \boldsymbol h_{\rm BR}^{los} + \sqrt{\frac{1}{1+\beta}} \boldsymbol h_{\rm BR}^{nlos}
\Bigg)}_{\text{array response \& small-scale fading}}; \label{eq.path1}\\
\boldsymbol h_{\rm JR}&=\underbrace{\sqrt{\rho d_{\rm JR}^{-\kappa_2}}}_{\text{path loss}}
\underbrace{\Bigg( \sqrt{\frac{\beta}{1+\beta}} \boldsymbol h_{\rm JR}^{los} + \sqrt{\frac{1}{1+\beta}} \boldsymbol h_{\rm JR}^{nlos}
\Bigg)}_{\text{array response \& small-scale fading}}; \label{eq.path2}\\
\boldsymbol h_{\rm RU}^t &=\sqrt{\rho (d_{\rm RU}^t)^{-2}} \boldsymbol g_{\rm RU}^t, ~\forall t. \label{eq.hrut}
\end{align}
\end{subequations}
Here, $\beta$ is the Rician factor of the B-R and J-R links (c.f. $\beta_t$); and
$\boldsymbol h_{\rm BR}^{los}$ and $\boldsymbol h_{\rm JR}^{los}$ are the LoS components, as given by
\begin{subequations}
\begin{align}
\boldsymbol h_{\rm BR}^{los}= & \left[1, \ldots, e^{-j\frac{2 \pi d_x}{\lambda}(N_x-1)\phi_{\rm BR}^x}\right]^T \otimes \notag \\
&\left[1, \ldots, e^{-j\frac{2 \pi d_y}{\lambda}(N_y-1)\phi_{\rm BR}^y}\right]^T, \\
\boldsymbol h_{\rm JR}^{los}= & \left[1, \ldots, e^{-j\frac{2 \pi d_x}{\lambda}(N_x-1)\phi_{\rm JR}^x}\right]^T \otimes  \notag \\
&\left[1, \ldots, e^{-j\frac{2 \pi d_y}{\lambda}(N_y-1)\phi_{\rm JR}^y}\right]^T,
\end{align}
\end{subequations}
%
where $d_x$ and $d_y$ are the antenna spacings in the directions of the $x$- and $y$-axes, respectively;
$\phi_{\rm BR}^x=x_{\rm R}/d_{\rm BR}$ and $\phi_{\rm BR}^y=y_{\rm R}/d_{\rm BR}$ are the spatial frequencies corresponding
to the angles-of-arrivals (AoAs) from BS to RIS,
and $\phi_{\rm JR}^x=(x_{\rm J}-x_{\rm R})/d_{\rm JR}$ and $\phi_{\rm JR}^y=(y_{\rm J}-y_{\rm R})/d_{\rm JR}$ are the spatial frequencies
corresponding to the AoAs from the jammer to the RIS along the $x$- and $y$-axes, respectively~\cite{zeng21}.

In \eqref{eq.path1} and \eqref{eq.path2}, $\boldsymbol h_{\rm BR}^{nlos} \in {\mathbb C}^{N \times 1}$
and $\boldsymbol h_{\rm JR}^{nlos} \in {\mathbb C}^{N \times 1}$ are the non-LoS (NLoS) components with the variables
independently drawn from the zero-mean, unit-variance CSCG distribution.
In \eqref{eq.hrut}, $\boldsymbol g_{\rm RU}^t$ is the array response, as given by
\begin{equation}\label{eq.grut}
\begin{aligned}
\boldsymbol g_{\rm RU}^t =& \left[1, \ldots, e^{-j\frac{2 \pi d_x}{\lambda}(N_x-1)\phi_{{\rm RU},x}^t}\right]^T \otimes \\
& \left[1, \ldots, e^{-j\frac{2 \pi d_y}{\lambda}(N_y-1)\phi_{{\rm RU},y}^t}\right]^T,~\forall t,
\end{aligned}
\end{equation}
where $\phi_{{\rm RU},x}^t=(x_t-x_{\rm R})/d_{\rm RU}^t$ and $\phi_{{\rm RU},y}^t=(y_t-y_{\rm R})/d_{\rm RU}^t$ are the spatial
frequencies corresponding to the angles-of-departures (AoDs) from RIS to UAV along the $x$- and $y$-axes, respectively.

It is noteworthy that the UAV does not ask for the CSI involving the RIS and the jammer to produce its flight path and configure the RIS in this paper.
Instead, the UAV only measures its own received data rate to learn
the control policy of its flight path and RIS configuration.
This consideration is of practical value, since the RIS-reflected channels are difficult and slow to estimate.

\section{Problem Formulation}\label{sec.prob}
Let $P_t, \forall t$ denote the transmit power of the BS and $P_{\rm J}$ denote the transmit power of the jammer.
The signal-to-interference-plus-noise ratio (SINR) at the UAV at time slot $t$, denoted by $\gamma_{\rm U}^t$,
is
\begin{equation}\label{eq.snrsnr}
\gamma_{\rm U}^t = \frac{P_t | h_{\rm BU}^t + (\boldsymbol h_{\rm RU}^t)^H \boldsymbol \Theta_t \boldsymbol h_{\rm BR} |^2}
{P_{\rm J} | h_{\rm JU}^t + (\boldsymbol h_{\rm RU}^t)^H \boldsymbol \Theta_t \boldsymbol h_{\rm JR} |^2+ \sigma_{\rm U}^2},
\end{equation}
where $\sigma_{\rm U}^2$ is the variance of the additive white Gaussian noise (AWGN) at the UAV.
The received data rate of the UAV at time slot $t$ is given by
\begin{equation}\label{eq.everate}
R_{\rm U}^t = \log_2 (1+\gamma_{\rm U}^t).
\end{equation}

We aim to maximize the total received data rate of the UAV from the BS over the mission duration of $T_w$ slots.
The problem considered is stated as follows.
\begin{subequations}\label{eq.prob1}
\begin{align}
&\max_{ \{\boldsymbol q_t, \boldsymbol \Theta_t, \forall t \} } \sum_{t=1}^{T_w} R_{\rm U}^t   \\
&\text{s.t.} ~ -\pi \le \theta_n^t < \pi, ~\forall n, t, \\
&\quad \;~  \eqref{eq.mobfix1}-\eqref{eq.mobfix6}. \notag
\end{align}
\end{subequations}
Problem \eqref{eq.prob1} is challenging for traditional convex solvers due to several reasons: First of all, the received data rate is a non-convex function of the UAV's flight path ${\boldsymbol q_t, \forall t}$ and the RIS phase shifts ${\boldsymbol \Theta_t, \forall t }$. Second, the UAV flight path waypoints are embedded in the exponents of the R-U link in \eqref{eq.hrut} and \eqref{eq.grut}, making the trajectory optimization intractable for existing convex tools, such as successive convex approximation.
Another reason is that the large number of RIS reflecting units can cause prohibitive overhead and complexity for radio channel estimation, acquisition and reconfiguration. 
To overcome these limitations, the next section proposes using DRL to solve~\eqref{eq.prob1}.

\section{Proposed DRL Framework for Anti-Jamming Communication of UAV-borne IoT platforms}\label{sec.ddpg}
The proposed method in this section aims to solve problem \eqref{eq.prob1} by utilizing the DDPG and TD3 models. Our approach involves learning to adjust the RIS and control the UAV's trajectory, including heading and acceleration, based on changes in the received data rate of the UAV. Importantly, our method eliminates the need for precise CSI or knowledge of the RIS reflecting channels. DDPG and its variations, such as TD3, have been demonstrated to be effective in addressing problems with continuous action spaces~\cite{Silver2014ddpg, Timothy2016Continuous, hu21dml}. In contrast, traditional DRL methods, such as deep Q-learning, can struggle and even diverge when faced with continuous action spaces.

\subsection{State, Action, and Reward}
Since the current UAV location only depends on its previous location and speed,
the UAV trajectory (i.e., waypoints) is a Markov decision process (MDP).
The RIS configuration depends solely on the instantaneous position of the UAV.
Therefore, we interpret problem \eqref{eq.prob1} as an MDP with its state, action, and reward defined below.
\begin{itemize}
	\item {\textbf{\emph{State Space ${\cal S}$: }}}At time slot $t$, the system state $s_t \in {\cal S}$ is made of the relative position of the UAV with regards to its final location, $ \boldsymbol q_t-\boldsymbol q_F$,
	the velocity of the UAV, $\boldsymbol V_t$, and the SINR at the UAV, $\gamma_{\rm U}^t$,
	$s_t=\{\boldsymbol q_t-\boldsymbol q_F, V_t, \gamma_{\rm U}^t\}$.
	
	\item {\textbf{\emph{Action Space $\cal A$: }}} It gathers all possible actions, i.e., $ a_t \in \mathcal A$. 
	During the $t$-th time step, the action $a_t$ consists of the reflecting coefficients $\{\theta^t_{n}\}_{n \in {\cal N}}$ and the acceleration of the UAV,
	$\boldsymbol A_t:=[A_{xt}, A_{yt}, A_{zt}]^T$, i.e., 
	$a_t = \left\lbrace \theta^{t}_{n} \in [-\pi, \pi), \forall n, A_{it} \in [-A_{\max},\,A_{\max}], \; i \in \{x,y,z\}  \right\rbrace$.
	The UAV acceleration is constrained by \eqref{eq.mobfix4}--\eqref{eq.mobfix6}.	
	Given the initial location and velocity of the UAV, its future waypoints $\boldsymbol q_t$ and velocities $\boldsymbol V_t$ 
are decided by the accelerations, i.e., by \eqref{eq.mobfix1}--\eqref{eq.mobfix3}.

	\item {\textbf{\emph{Reward $r_t$: }}}The reward function gives positive returns per time step for implementing action $a_t$: 
	\begin{align}\label{eq-reward}
	r_t = \underbrace{R_{\rm U}^t}_{\text{communication}}
	 + \underbrace{\zeta \left(d^{t-1}_F - d^t_F \right)}_{\text{distance to the final location}},
	\end{align}	
	where $d^{t-1}_F = \|\boldsymbol q_{t-1}-\boldsymbol q_F \|$ and $d^{t}_F = \|\boldsymbol q_{t}-\boldsymbol q_F \|$
	are the distances from the UAV to the final location at the $(t-1)$-th and $t$-th time steps, respectively;
	and $\zeta$ is a tunable parameter during the learning process.
	The second element on the right-hand side of \eqref{eq-reward} encourages the UAV to fly towards the final location. 
	
	
	\item {\textbf{\emph{Policy: }}}A projection from the state space, ${\cal S}$, to the action space ${\cal A}$ is referred to as a policy, $\mu: {\cal S} \to {\cal A}$, a distribution 
	${\mu}(a | s) = \Pr\left(a_t = a | s_t =s \right)$ 
	over state $s \in~{\cal S}$. 
	
	\item {\textbf{\emph{Experience: }}}The experience, defined as $e_t = \left( s_t,a_t,r_t,s_{t+1} \right) $, is stored in an experience replay memory~${\bm{R}}$.
\end{itemize}
%
 
The UAV experiences state $s_t$, performs action $a_t$, receives reward $r_t$, and turns to state $s_{t+1}$.
A policy $a_t = \mu(s_t)$ maps state $s_t$ to a possible action. The UAV chooses the policy that maximizes the cumulative reward $R_t = \sum_{n=t}^N \gamma^{n-t} r_t$. Here, $\gamma \in (0,1)$ gives the discount factor. Given $s_t$, $a_t$, and $\mu$, the Q-function evaluates $R_t$ by 
\begin{equation}
Q_{\mu}(s_t, a_t)=\mathbb E_{\mu}[R_t | s_t, a_t].
\end{equation}
The action-value function, $Q_{\mu}(s_t, a_t)$, follows the Bellman Expectation Equation:
\begin{equation}\label{eq-bellman}
\begin{aligned}
Q_{\mu}\left(s_t,a_t \right) & = \mathbb{E}_{r_t, s_{t+1} \sim {\rm\cal E} }\left[r_t + \gamma \mathbb{E}_{a_{t+1}\sim \mu} \left[Q_{\mu}\left(s_{t+1},a_{t+1} \right) \right] \right].
\end{aligned}	
\end{equation}
Here, ${\cal E}$ stands for the environment the UAV experiences.

It is generally challenging to directly use an RL algorithm to solve the continuous-space, finite-horizon MDP and determine the Q-value, $Q(s_t, a_t)$, due to the continuous state and action spaces. This paper puts forth a new DDPG-based algorithm to control the UAV's trajectory and configure the RIS, as delineated in the following subsection.

\subsection{Actor-Critic Framework-Based DDPG}
\begin{figure*}[tb]
	\centering
	\includegraphics[width=1.8\columnwidth]{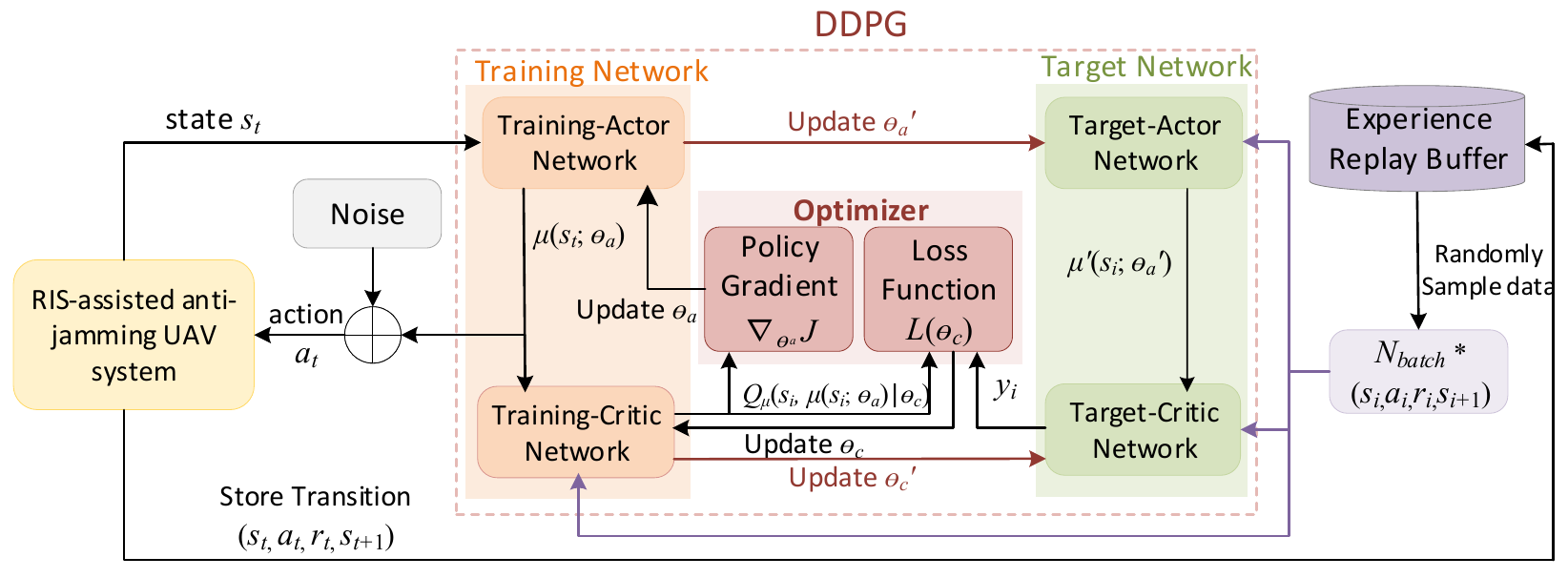}
	\caption{The proposed DDPG-based framework with a training network and a target network, each comprising an actor network and a critic network. The experience replay buffer gives batches of samples of state transitions for training and updating the networks.}
	\label{fig-actor-critic}
\end{figure*}
\begin{figure*}[tb]
	\centering
	\includegraphics[width=1.8\columnwidth]{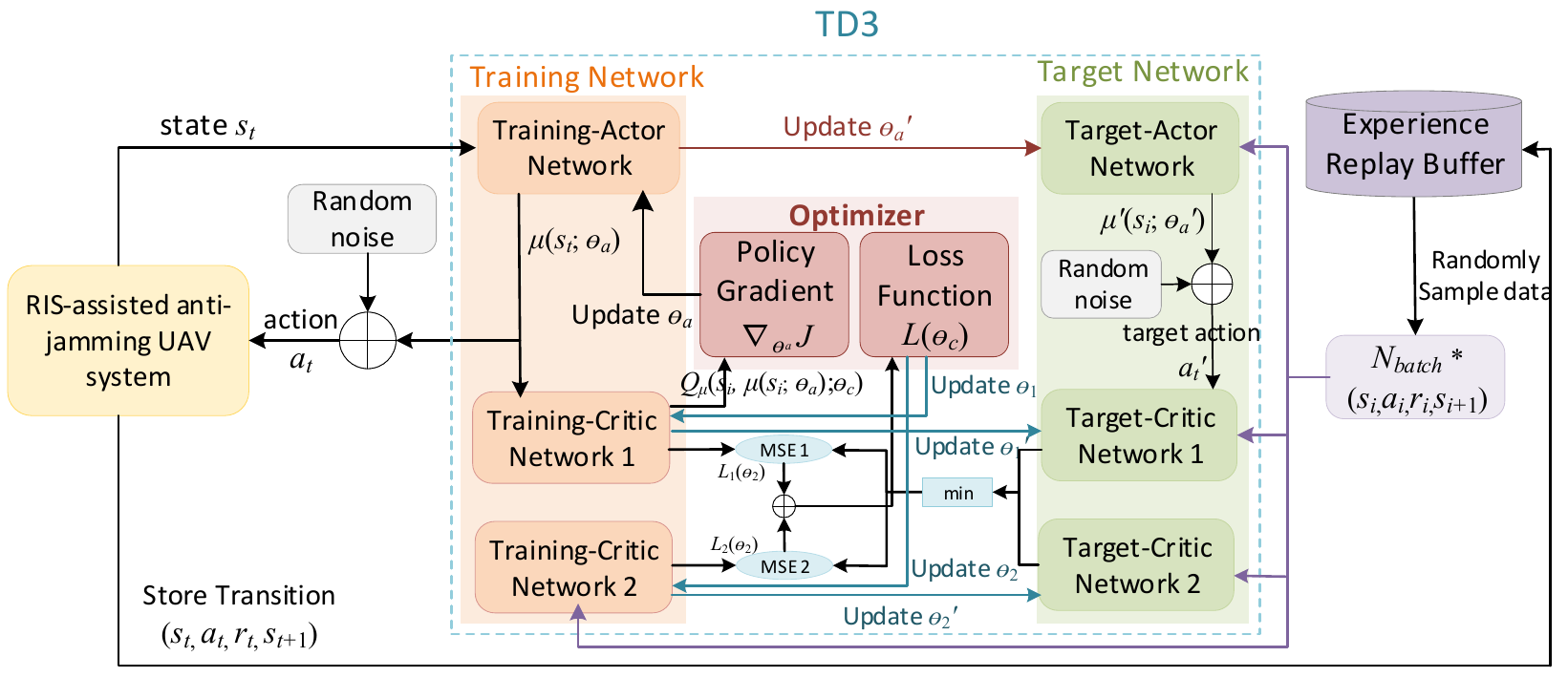}
	\caption{The proposed TD3-based framework with an actor network comprising an actor and a target-actor, and a critic network comprising two critics and two target-critics. The experience replay buffer gives batches of samples of state transitions for training and updating the networks.}
	\label{fig-td3}
\end{figure*}



The DDPG-based network uses four DNN approximators, including training-actor and training-critic networks, and target-actor and target-critic networks, as shown in Fig.~\ref{fig-actor-critic}. The training-actor network with parameters $\theta_a$, denoted as $\mu\left(s_t; \theta_a \right)$, gives an approximate policy of the UAV and produces the actions. The training-critic network with parameters $\theta_c$, denoted as $Q_{\mu}(s_t,a_t;\theta_c)$, estimates the action-value function concerning the actions created in the training-actor network~\cite{Silver2014ddpg}. The target-actor network with parameter $\theta'_a$, represented by $\mu'\left(s_t; \theta'_a \right)$, and the target-critic networks with parameter $\theta'_c$, represented by $Q'_{\mu'}(s_t,a_t;\theta'_c)$, generate the target Q-value for training the training-actor and training-critic networks.

The DDPG network uses the deterministic policy gradient (DPG) theorem~\cite{Silver2014ddpg} to refresh $\theta_a$, $\theta_c$, $\theta'_a$, and $\theta'_c$. It produces actions in an actor-critic setting. Additionally, the adoption of a target network (i.e., the target-actor and target-critic networks) helps prevent unstable learning, as opposed to using only a training network (with a training-actor and a training-critic network)~\cite{Hou2017}.

The UAV inputs state $s_t$ into the training-actor network. Using the DPG theorem~\cite{Silver2014ddpg}, the network generates the strategy by projecting the state to an action in a deterministic fashion. This network approximates the agent's policy function and selects action $a_t$. A noise is added to $a_t$ to balance between new and known actions, resulting in an output action $a_t = \mu\left(s_t; \theta_a \right) + {\cal N}_t$. Herein, ${\cal N}_t$ is a random noise process with a normal distribution. The agent is rewarded with $r_t$ and transitions to state $s_{t+1}$. Then, it stores the experience $(s_t, a_t, r_t, s{t+1})$ in ${\bm R}$.

The training-critic network evaluates the action-value function $Q_{\mu}\left(s_t, \mu(s_t;\theta_a);\theta_c \right)$ of the selected action $a_t$. By using a random sample from the replay memory $\bm R$, the network approximates the action-value function as $Q_{\mu}(s_i, \mu(s_i;\theta_a);\theta_c)$.
We take $J(\theta_a)$ to be the probability distribution of the parameter $\theta_a$. The training-actor network is adjusted in the direction that improves the strategy the most rapidly, i.e., in the direction of the gradient of $J(\theta_a)$ with respect to (w.r.t.) $\theta_a$ 
\cite{Silver2014ddpg}: 
\begin{subequations}\label{eq-gradient}
	\begin{align}
	&\nabla_{\theta_a} J(\theta_a) = \mathbb{E}_{s \sim \rho^{\mu}} \left[\nabla_{\theta_a} {{Q}}_{\mu}(s_t,\mu(s_t;\theta_a);\theta_c) \right]\label{eq-gradient a}\\
	&\;\; =  \mathbb{E}_{s \sim \rho^{\mu}} \left[\nabla_{\theta_a} \mu(s_t;\theta_a) \nabla_{a} {{Q}}_{\mu}(s_t,a;\theta_c)|_{a = \mu(s_t;\theta_a)} \right],\label{eq-gradient b}
	\end{align}
\end{subequations}
where \eqref{eq-gradient b} uses the chain rule; 
$\rho^{\mu}$ provides a discounted state distribution of  $\mu(s_t;\theta_a)$~\cite{Timothy2016Continuous};
$\nabla_{\theta_a} \mu(s)$ gives the gradient of the training-actor network $\mu(s)$ w.r.t. $\theta_a$; $\nabla_{a} {{Q}}_{\mu}(s_t,a;\theta_a)$ provides the gradient of ${{Q}}_{\mu}(s_t,a;\theta_a)$ w.r.t.~$a$. 

By randomly drawing $N_{batch}$ sampled historical transitions from ${\bm R}$, the gradient $\nabla_{\theta_a} J(\theta_a)$ is approximated by
\begin{equation}\label{eq-gradient-approx}
\nabla_{\theta_a} J(\theta_a) 
\!\approx\! \frac{1}{N_{batch}} \sum_{i = 1}^{N_{batch}}\!\left[\!\nabla_{\theta_a} \mu(s_i) \nabla_{a} {{Q}}_{\mu}(s_i,a;\theta_c)|_{a = \mu(s_i)} \!\right].
\end{equation}
The training-actor network  parameter, i.e., $\theta_a$, is refreshed based on the gradient ascent~\cite{sutton1999policy}
\begin{equation}\label{eq-policy-ascent}
\begin{aligned}
\theta_a & \leftarrow \theta_a + \eta_a \nabla_{\theta_a} J(\theta_a)\\
& \approx \theta_a + \frac{\eta_a}{N_{batch}} \!\sum_{i = 1}^{N_{batch}}\!\left[\nabla_{\theta_a} \mu(s_i) \nabla_{a} {{Q}}_{\mu}(s_i,a;\theta_c)|_{a = \mu(s_i)} \!\right].
\end{aligned}
\end{equation} 
Here, $\eta_a$ specifies the learning rate of the training-actor network.

The training-critic network is refreshed through minimizing the following loss function:
\begin{equation}\label{eq-loss 1}
\begin{aligned}
L(\theta_c) & = \mathbb{E}_{s_t \sim \rho^{\mu}, a_t \sim \mu(s_t;{\theta_a})}\left[\left( Q_{\mu}\left(s_t, a_t;\theta_c \right)\! -\! y_t \right)^2\right].
\end{aligned}
\end{equation}
Here, $y_t = r_t + \gamma Q'_{\mu'}\left(s_{t+1},\mu'\left( s_{t+1};\theta'_a\right); \theta'_c\right)$ is the target Q-value produced by the target network under the transition $(s_t, a_t,r_t, s_{t+1})$.
Here, the parameters of the target-actor and
target-critic networks, $\theta'_a$ and $\theta'_c$, are the respective decayed copies of $\theta_a$ and $\theta_c$.

With $N_{batch}$ randomly sampled transitions, the loss function, $L(\theta_c)$, is approximately evaluated by
\begin{equation}\label{eq-loss 2}
\begin{aligned}
L(\theta_c) \approx \frac{1}{N_{batch}} \sum_{i = 1}^{N_{batch}}\left[\left( Q_{\mu}\left(s_i, \mu(s_i; \theta_a);\theta_c \right) - y_i \right)^2\right],
\end{aligned}
\end{equation}
where $y_i = r_i + \gamma Q'_{\mu'}\left(s_{i+1},\mu'\left( s_{i+1};\theta'_a\right) ; \theta'_c\right)$ gives the approximate target Q-value that the target network generates upon $N_{batch}$ transitions sampled at random. 
By differentiating $L(\theta_c)$ w.r.t. $\theta_c$, the gradient is attained: 
\begin{equation}
\begin{aligned}
\nabla_{\theta_c} L(\theta_c) 
&\!\approx\! \frac{2}{N_{batch}} \!\sum_{i = 1}^{N_{batch}}\!\left[\left( Q_{\mu}\left(s_i, \mu(s_i; \theta_a);\theta_c \right) - y_i \right) \right. \\
&\qquad \qquad \qquad \left.\times \nabla_{\theta_c} {{Q}}_{\mu}(s_i,\mu(s_i; \theta_a);\theta_c)\right]. 
\end{aligned}	
\end{equation}
The training-critic network parameter, $\theta_c$, is refreshed by utilizing the stochastic gradient descent method~\cite{sutton1999policy}. 

The target-actor and target-critic networks are refreshed based on the training-actor and training-critic networks: 
\begin{equation}
\begin{aligned}
\theta'_a & \leftarrow \tau_a \theta_a + (1-\tau_a)\theta'_a,\\
\theta'_c & \leftarrow \tau_c \theta_c + (1-\tau_c)\theta'_c,
\end{aligned}	
\end{equation} 
where $\tau_a$ and $\tau_c$ are the decaying rates for the training-actor and training-critic networks, respectively. 

\subsection{Twin Delayed DDPG (TD3)}
TD3 is one of the latest extensions of DDPG and consists of a training network and a target network, where the training network is made of a
training-actor and two training-critic networks, and the target network comprises a target-actor and two target-critic networks, as shown in Fig.~\ref{fig-td3}.
TD3 addresses the Q-value overestimation problem of the DDPG algorithm by incorporating \emph{three improvements} over the classical DDPG model, namely, \emph{clipped double-Q learning}, \emph{target policy smoothing}, and
\emph{delayed policy update}~\cite{dankwa2019twin, yuan23}.

\begin{itemize}
	\item {\emph{Clipped double-Q learning: }} TD3 contains two training-critic and target-critic networks to produce two Q-values. The lesser of the two is used to evaluate the target Q-value in the Bellman error loss function. Specifically, $Q_{\mu}(s_t, a_t; \theta_c)$ in the DDPG is replaced by $Q_{\mu}(s_t, a_t) = \min \left\{Q_{1}\left(s_t, a_t;\theta_1 \right), Q_{2}\left(s_t, a_t;\theta_2 \right) \right\}$ in the TD3.
	
	\item {\emph{Target policy smoothing: }}TD3 perturbs actions produced by the target-actor network (i.e., ``target action'') with noises and smooths the corresponding Q-function values to enhance the resistance of the policy against erroneous Q-functions.
	The smoothed target action is written as
	\begin{equation}
	a'_{t} = {\text{clip}}\left( \mu'\left(s_{t+1}; \theta'_a \right) + {\text{clip}\left( \epsilon', -\sigma_m^2, \sigma_m^2\right)},
	a_{\min}, a_{\max}\right).
	\end{equation}
	Here, the noise $\epsilon'$ is taken at random from a Gaussian distribution 
	with zero mean and variance $\sigma_a^2$, i.e., $\epsilon' \sim {\cal N}(0,\sigma_a^2)$; 
	and $\sigma_m^2$ is the maximum exploration noise supported by the environment. 	In contrast, the DDPG model does not add noises towards target actions.
	
	\item {\emph{``Delayed'' policy update: }}The training-actor and target-actor networks (i.e., policies) are refreshed less frequently than the training-critic and target-critic networks.
	For example, it was recommended in \cite{dankwa2019twin} that the training-actor and target-actor networks are refreshed after the training-critic and target-critic networks are refreshed twice in TD3.
	In contrast, the classical DDPG model refreshes its train-actor and target-actor networks and train-critic and target-critic networks at the same pace. 
\end{itemize}


\section{Performance Evaluation}\label{sec.sim}
We carry out extensive experiments in Python to evaluate the proposed approach. The location of the jammer is $\boldsymbol q_{\rm J}=[-25, -25, 0]^T$~m.
The UAV's initial and final locations are $\boldsymbol q_0=[-200, -100, 5]^T$~m
and $\boldsymbol q_F=[100, 60, 50]^T$~m.
The RIS has $N=5 \times 4=20$ (or $N=5 \times 8=40$) reflecting elements,
and the reference point is $\boldsymbol q_{\rm R}= [50, 50, 30]^T$~m.
The scheduling horizon is $T=30$ s with each time slot being $\delta=0.1$ s.
The other parameters concerning the system model are collated in Table~\ref{tab.para}.

\begin{table}[t]
		\renewcommand{\arraystretch}{1.0}
		\caption{The Parameters of the System Model}
		\begin{center}
			\begin{tabular}{ll}
				\toprule[1.0pt]
				Parameter  & Value \\
				\hline
RIS antenna separation, $d_x$, $d_y$   & $\lambda/2$   \\ 
Maximum and minimum speeds, $V_{\max},~V_{\min}$                &40 m/s, 2 m/s \\ 
Path loss, $\rho$         & -30 dB \\ 
Path loss exponents, $\kappa_1,~\kappa_2$          & 3.5, 2.8 \\ 
Rician factor coefficients, $\xi_1,~\xi_2$     & 1, 4.4   \\ 
Rician factor, $\beta$                               & 3 dB \\ 
Noise power, $\sigma_{\rm U}^2$                          &-169 dBm \\ 
				\toprule[1.0pt]
			\end{tabular}
		\end{center} \label{tab.para}	
\end{table}

\subsection{Experiment Settings}
The proposed DDPG network is composed of actor networks implemented using fully connected neural networks (FCNNs) with three hidden layers and learning rates of $10^{-4}$. The first, second, and third layers of the actor networks have 64, 128, and 64 neurons, respectively. The output layer implements the $\tanh(\cdot)$ activation function to bound the output actions within $[-\pi, \pi)$ for the RIS configuration and $[-2, 2]$~m/s$^2$ for the UAV control. Additionally, the paper utilizes critic networks that employ FCNNs with two hidden layers and learning rates of $10^{-3}$. Both hidden layers utilize the Rectified Linear Unit (ReLU) activation functions with 64 neurons in the first layer and 128 neurons in the second layer. The DDPG actor policy is trained using additive noise $\cal N$, which is sampled from a complex Gaussian noise distribution with zero mean and variance 0.2.

The proposed TD3 network is built upon the DDPG network. It includes two duplicates of the training-critic and target-critic networks; see Fig. 3. Similar to the DDPG network, the actor in the TD3 network is trained using exploration noise that is drawn from a complex Gaussian distribution with zero mean and variance 0.2. Additionally, the target-actor in the TD3 network is smoothed using policy noise that is drawn from a complex Gaussian distribution with zero mean and variance 0.2. The maximum exploration noise is set to 0.5, and the actor networks are refreshed every two steps. The TD3 improves the DDPG by providing faster and smoother convergence, which is especially beneficial for larger RISs.

\begin{table}[t]
	\renewcommand{\arraystretch}{1.0}
	\caption{The hyperparameters of the proposed DDPG and TD3 algorithms}
	\begin{center}
		\begin{tabular}{ll}
			\toprule[1.0pt]
			Parameter  & Value \\
			\hline
			Reduction coefficient for upcoming reward, $\gamma$ & 0.99 \\
			Training coefficient for actor and critic networks, $\eta_a$, $\eta_c$  & $1 \times 10^{-4}$ \\
			Declining coefficient for actor and critic networks, $\tau_a$, $\tau_c$, $\rho_{\tau}$ & $5 \times 10^{-3}$\\
			Capacity for experience repetition & $1 \times 10^5$\\
			Quantity of episodes, $T_{ep}$ & 3000 \\
			Total steps per episodes, $T_s$ & 300 \\
			Quantity of experiences in a mini-batch, $N_{batch}$ & 128\\	
			Variance of the exploration noise, $\sigma_e^2$ & 0.2 \\	
			Delayed policy update interval (TD3) & 2	\\	 
			Variance of the policy noise (TD3), $\sigma_a^2$ & 0.2 \\
			Largest value of the Gaussian noise (TD3), $\sigma_m^2$ & 0.5 \\			
			\toprule[1.0pt]
		\end{tabular}
	\end{center} \label{table_hyper_td3}	
\end{table}

The hyperparameters of the proposed DDPG and TD3 networks are summarized in Table~\ref{table_hyper_td3}.
The DDPG and TD3 networks are trained on a server equipped with a NVIDIA Tesla P100 SXM2 16GB GPU. 

{\textbf{\emph{Baseline~1:}}} This baseline applies TD3 to UAV flight path planning in the absence of the RIS, referred to as ``without RIS''.
The TD3 algorithm learns the UAV's trajectory solely based on the received data rate at the UAV without CSI involving the RIS or jammer.

{\textbf{\emph{Baseline 2:}}}
This baseline decouples the UAV's trajectory plan from the RIS configuration by first using a TD3-based algorithm to optimize the UAV's trajectory given the RIS configuration, and then maximizing the signal-to-noise ratio (SNR) at each time slot using the Dinkelbach method under the assumption of perfect and instantaneous CSI for all involved channels.
The Dinkelbach method is used to reformulate the SNR maximization problem as a fractional program defined as $F(\gamma_{\rm U}^t)=\min_{\Theta_t}~ f(\boldsymbol \Theta_t ) - \gamma_{\rm U}^t g(\boldsymbol \Theta_t )$, s.t. $-\pi \le \theta_n^t < \pi, \forall n, t$, where $\gamma_{\rm U}^t = \frac{f(\boldsymbol \Theta_t )}{g(\boldsymbol \Theta_t)}$ and $\boldsymbol{\Theta}_t$ gives the RIS configuration. Given $\gamma_{\rm U}^t$, the fractional program can be reorganized as a quadratic program with a unit-modulus constraint and solved using manifold optimization. The value of $\gamma_{\rm U}^t$ is refreshed based on the resultant $\boldsymbol{\Theta}_t$. This process is repeated until convergence, and the convergent value of $\gamma_{\rm U}^t$ is output~\cite{dinkelbach1967nonlinear}.

\subsection{Results of Policy Learning}\label{sec.train}
\begin{figure}[t]
\centering
\includegraphics[width=0.5\textwidth]{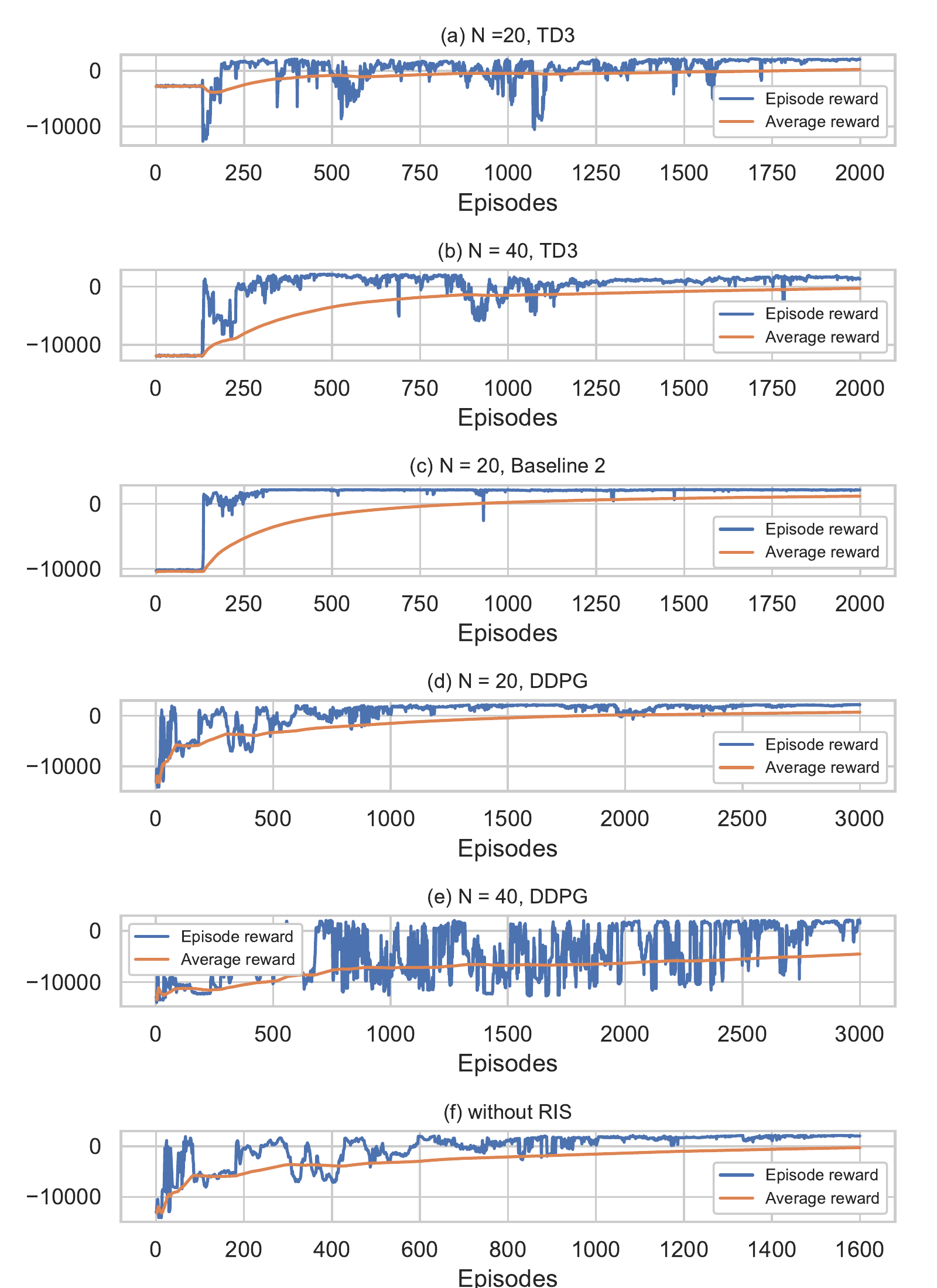}
\caption{The per-episode and average rewards of the proposed DDPG and TD3 algorithms.
Fig.~\ref{fig.reward_sub}(a) plots the proposed TD3 scheme when $N=20$; Fig.~\ref{fig.reward_sub}(b) plots the proposed TD3 scheme when $N=40$; Fig.~\ref{fig.reward_sub}(c) plots Baseline~2 when $N=20$; Fig.~\ref{fig.reward_sub}(d) plots the proposed DDPG scheme when $N=20$; Fig.~\ref{fig.reward_sub}(e) plots the proposed DDPG scheme when $N=40$;
and Fig.~\ref{fig.reward_sub}(f) plots Baseline~1 ``without RIS''.}
	\label{fig.reward_sub}
\end{figure}

Fig.~\ref{fig.reward_sub} plots the per-episode and average rewards of the proposed and baseline algorithms for~$N=20$ and~$N=40$.
The average reward for the $i$-th training episode, denoted by $\bar r_i$, is evaluated as 
$\bar r_i = \frac{1}{i}\sum_{j=1}^{i} r_j$,
where $i=1,\cdots,T_{ep}$, and $r_j$ is the step reward for the $j$-th training episode; see~\eqref{eq-reward}.

Fig.~\ref{fig.reward_sub} shows that the average reward gradually increases,
as the UAV control policy adapts to a randomly generated target trajectory in each episode.
The proposed TD3 algorithm outperforms the baselines,
and achieves its maximum reward at the 654th episode for $N =20$, and at the 477th episode for $N =40$.
The proposed DDPG algorithm reaches its maximum reward at the 1,607th episode when $N =20$, and at the 2,995th when $N =40$.
Baseline 1 without RIS reaches its maximum reward at the 280th episode.
Baseline 2 reaches its maximum reward at the 1,752nd episode when $N =20$.
The fast convergence of Baseline 2 is due to its substantially smaller action space of only UAV accelerations resulting from an unrealistic assumption of perfect and instantaneous CSI of all involved channels.
In general, the TD3 converges faster and more smoothly than the DDPG.
Yet, it undergoes less smooth changes in the per-episode reward when the action space is smaller, i.e., $N=20$,
This is because the TD3 has two critic networks for both the training and target networks (c.f. Fig.~\ref{fig-td3}),
which could incur higher complexity and more randomness when training, especially when the action space is smaller.
In contrast, DDPG is suitable for training tasks that are less complicated and  have relatively smaller action spaces.

\subsection{Test Results of Learned Policy}\label{sec.testing}
\begin{figure}[t]
	\centering
	\includegraphics[width=0.5\textwidth]{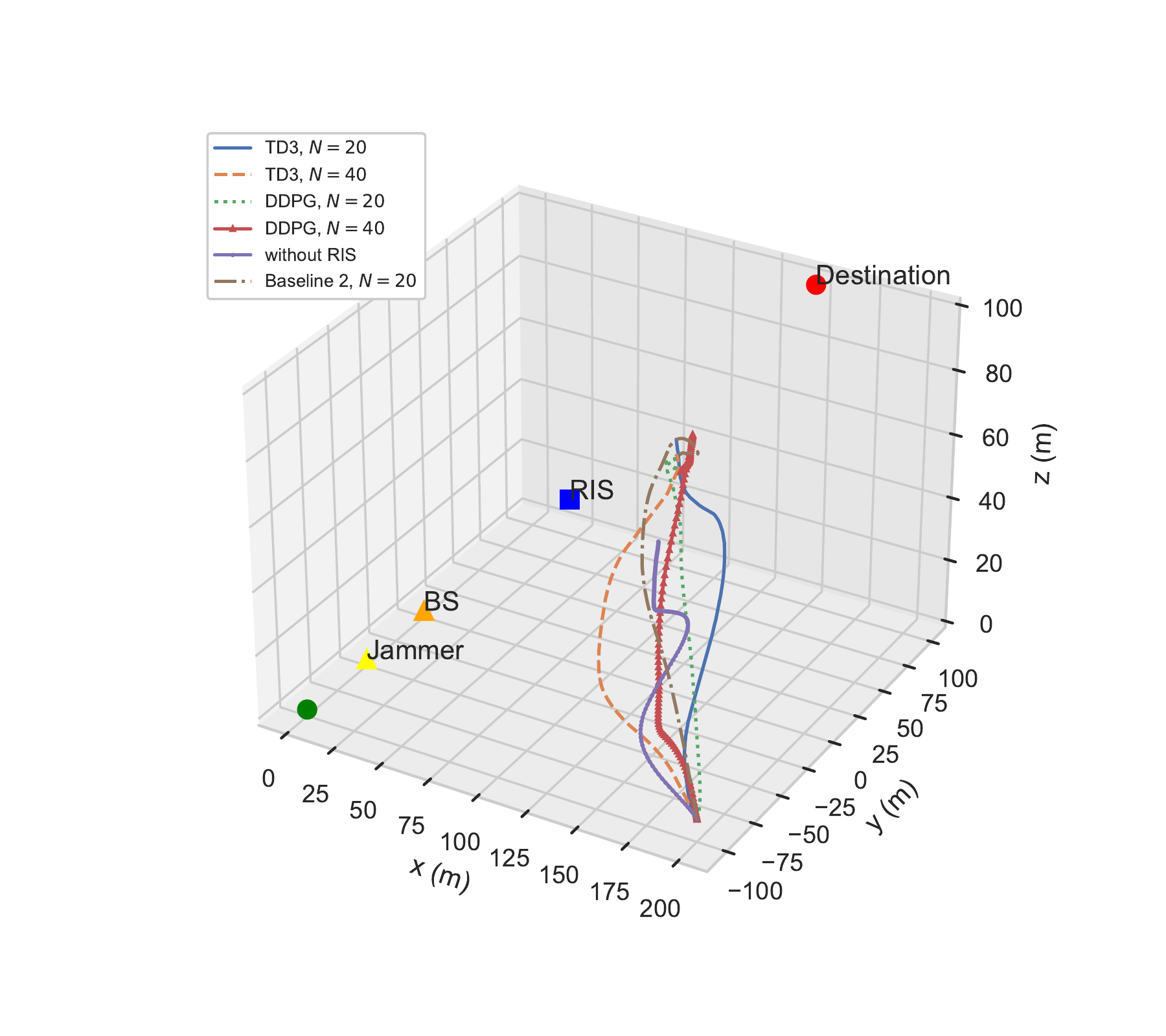}
	\caption{3D UAV trajectory, where the green and red dots are the UAV initial location and its expected destination, the orange and yellow triangles denote the locations of the BS and Jammer, and the blue square is the RIS reference point.}
	\label{fig.3D_trajectory}
\end{figure}

\begin{figure}[t]
	\centering
	\includegraphics[width=0.5\textwidth]{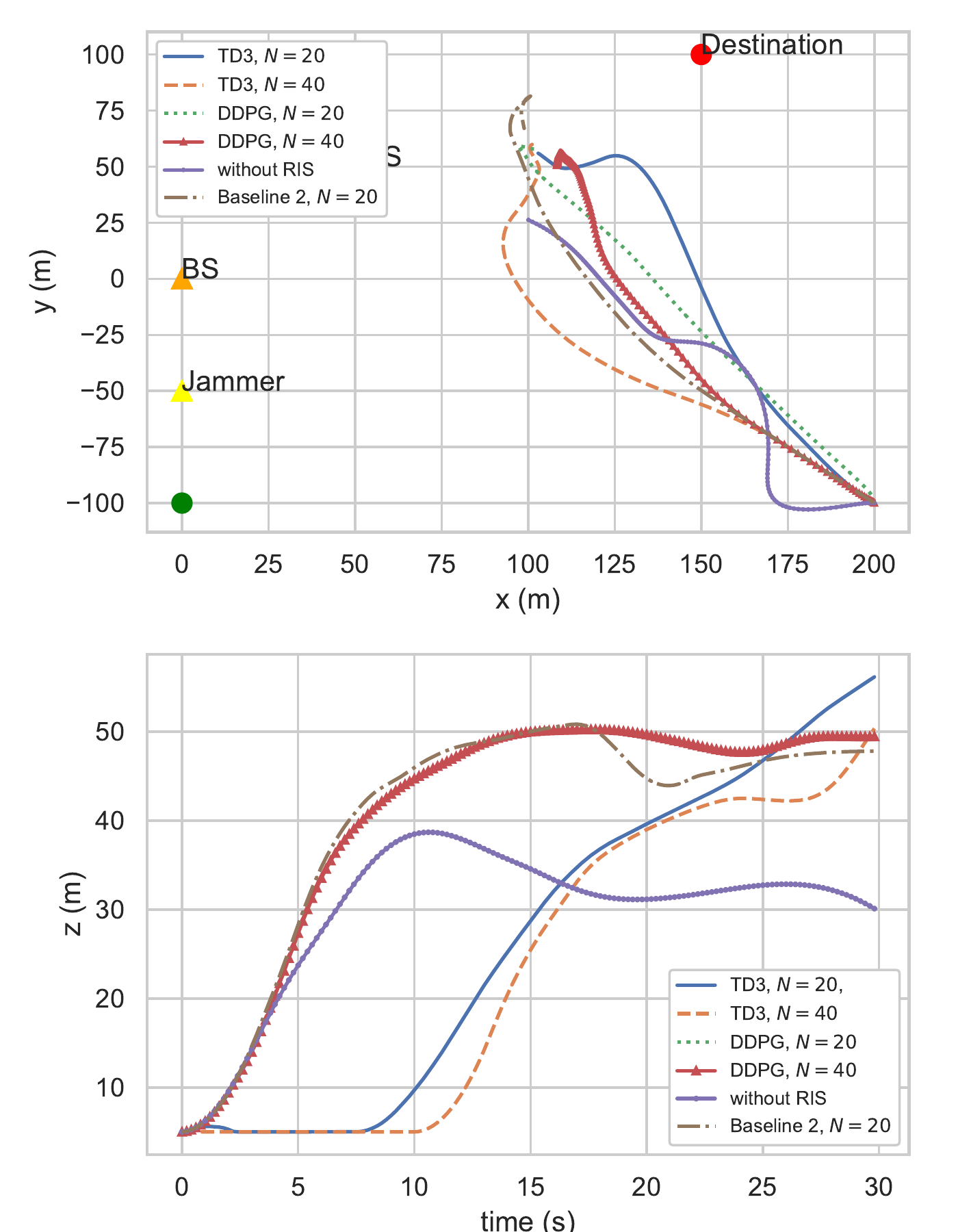}
	\caption{Projection of the UAV trajectory on the $x$-$y$ plane, with the green and red dots being the UAV initial and final locations, the orange and yellow triangles being the locations of the BS and Jammer, and the blue square being the RIS reference point.}
	\label{fig.2D_trajectory}
\end{figure}

Using the learning results obtained in Section~\ref{sec.train}, we test the proposed DDPG-based and TD3-based algorithms
for $N =20$ and $40$ at the RIS, as well as the baseline schemes for $N=20$ at the RIS for comparison. 
500 testing episodes are conducted, each consisting of 300 steps. During testing, no exploration noise is added.
The proposed algorithms and baselines are evaluated in terms of the 3D UAV trajectory and received data rate. Fig.~\ref{fig.3D_trajectory} shows the 3D trajectory of the UAV, while Fig.~\ref{fig.2D_trajectory} illustrates the trajectory in the $x$-$y$ plane and along the $z$-axis. These figures demonstrate that the UAV is able to adapt its control policy to the anti-jamming communication and successfully reach the destination.



\begin{figure}[t]
	\centering
	\includegraphics[width=0.5\textwidth]{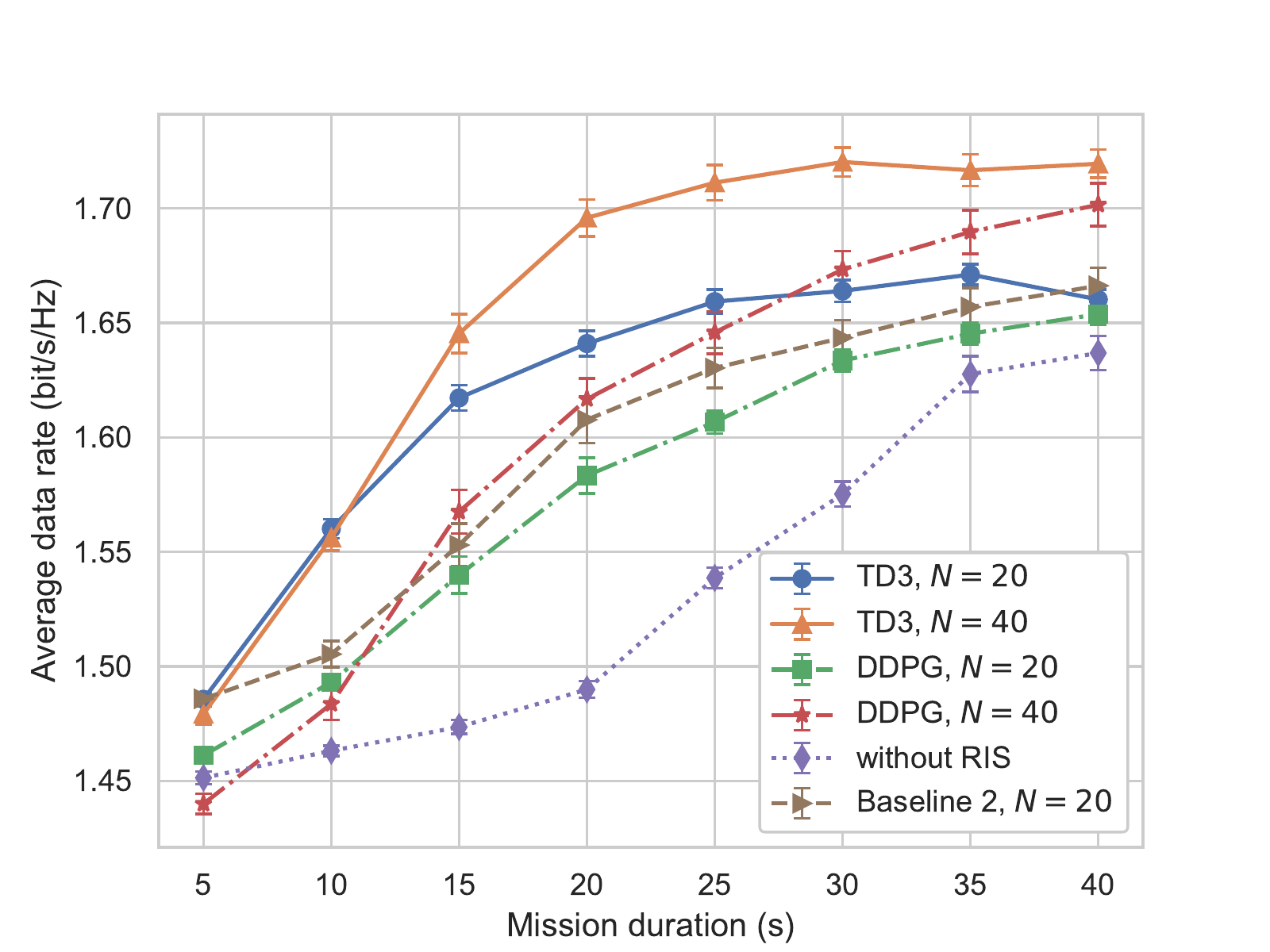}
	\caption{The received data rate of the UAV vs. the mission duration averaged over $500$ independent testing runs.}
	\label{fig.data_rate}
\end{figure}

Fig.~\ref{fig.data_rate} shows the received data rate of the UAV as the mission duration increases. The results are based on an average of 500 independent testing episodes, with error bars representing the associated uncertainty. It is observed that the received data rate increases with the mission duration under all considered algorithms, and the use of an RIS improves the received data rate. The proposed TD3 and DDPG algorithms give slightly lower data rates than Baseline 2, but operate without the CSI involving the RIS or the jammer, demonstrating their ability to adapt to system changes. Additionally, the proposed TD3 algorithm achieves a substantially higher data rate than the proposed DDPG algorithm, particularly for larger numbers of RIS elements. This is attributed to the faster convergence and better convergent control policy of TD3 compared to DDPG, as previously shown in Fig.~\ref{fig.reward_sub}.


\begin{figure}[th]
	\centering
	\includegraphics[width=0.5\textwidth]{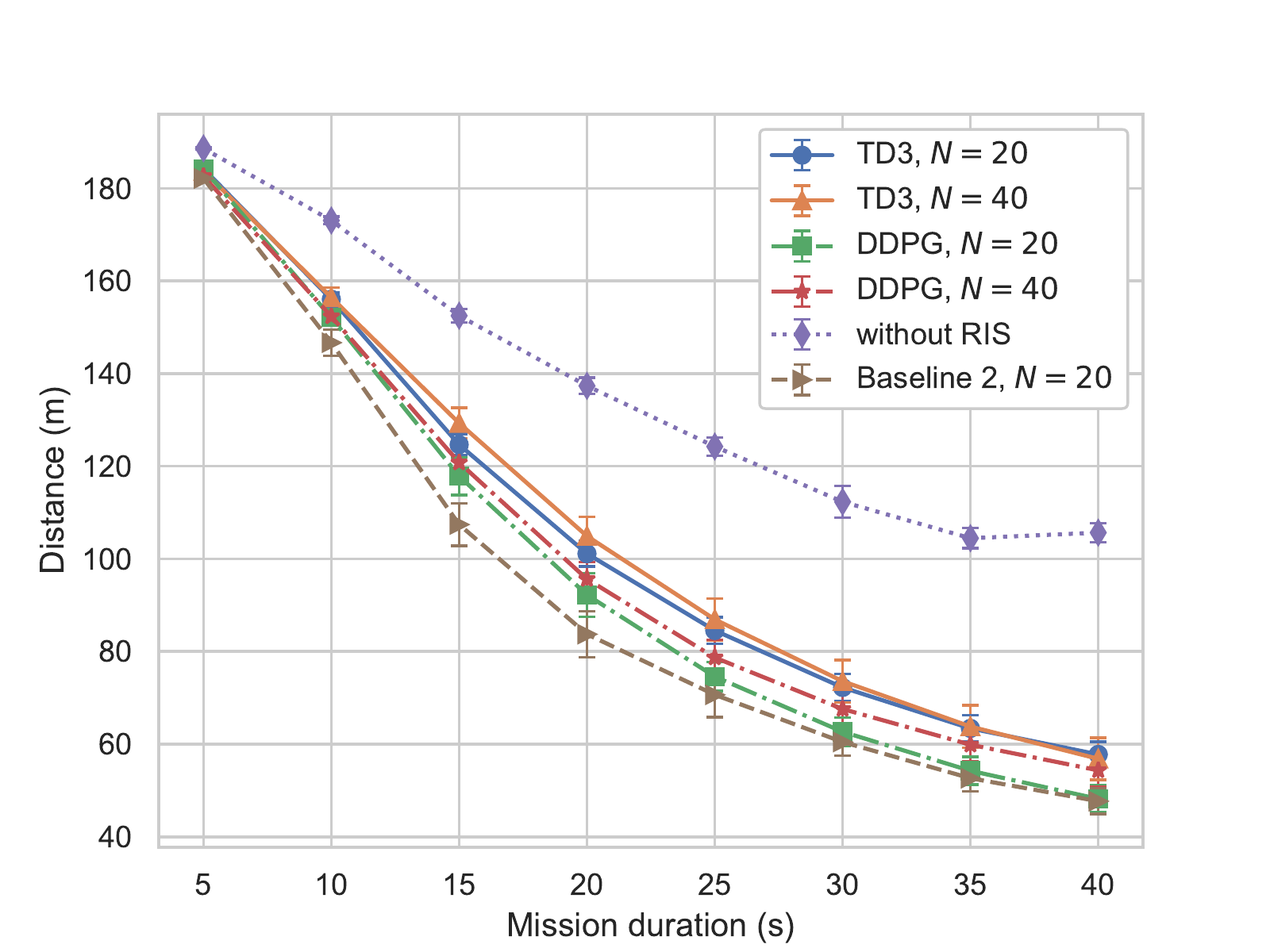}
	\caption{The finishing distance between the UAV and its destination at the end of the mission with the increase of the mission duration.}
	\label{fig.avg_distance}
\end{figure}
\begin{figure}[th]
	\centering
	\includegraphics[width=0.5\textwidth]{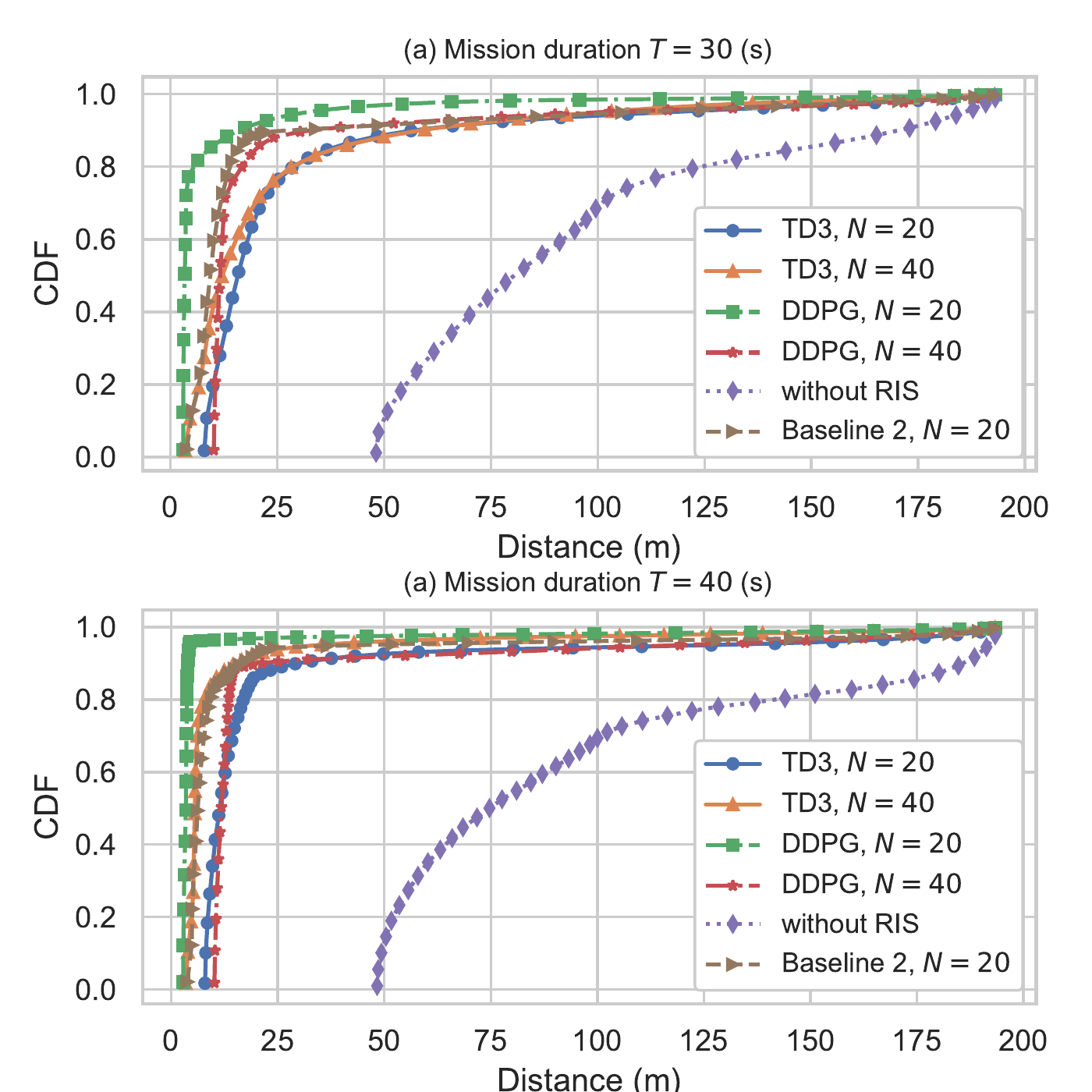}
	\caption{Cumulative distribution function (CDF) of the distance from the UAV to the final location
	when the mission duration is $T = 30$ and $40$ s.}
	\label{fig.cdf_dis}
\end{figure}


\begin{figure}[t]
	\centering
	\includegraphics[width=0.5\textwidth]{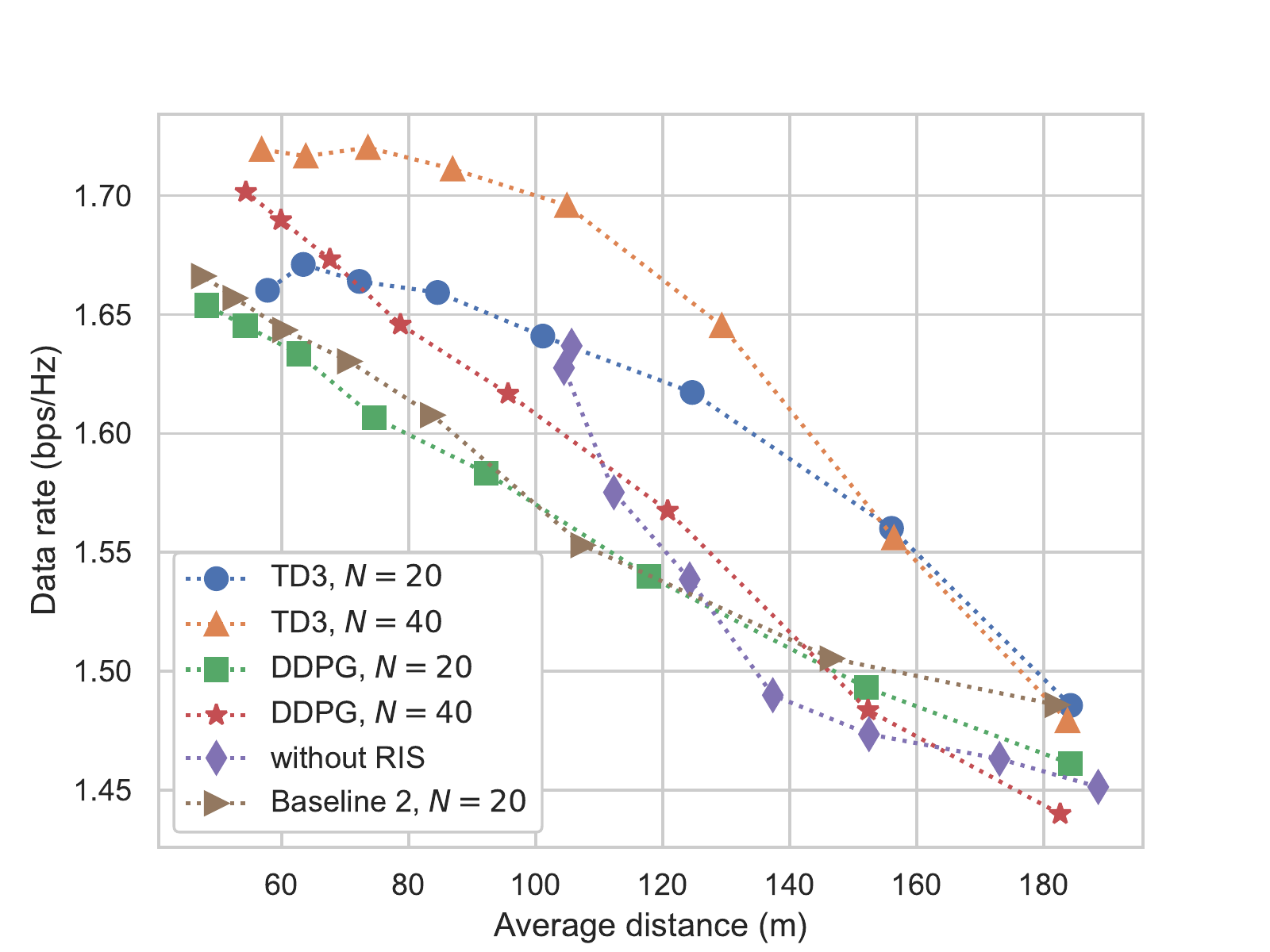}
	\caption{The UAV's data rate vs. its distance to the expected destination, where $T$ ranges from $5$ to $40$ seconds.}
	\label{fig.dis_rate}
\end{figure}

Fig.~\ref{fig.avg_distance} illustrates the distance between the UAV and its expected destination at the end of the mission. The results show that as the mission duration increases, the distance decreases for all algorithms. Without the use of the RIS, the UAV is unable to reach its expected final location as it must remain close to the BS to maintain a sufficient data rate. However, the proposed algorithms, such as TD3, enable the UAV to get closer or reach its final location by adjusting the RIS to enhance the desired signals, weaken the jamming signals and extend the effective BS-UAV transmission range.

It is worth noting that as the mission duration increases, DDPG can increasingly approach TD3 in the average achievable data rate. This indicates that DDPG is suitable for a less constrained problem setting where there is sufficient time for the UAV to maneuver and explore its action space.
In this case, DDPG can be a suitable solution, as it can benefit from its simpler network architecture than TD3.
In contrast, TD3 demonstrates significant gains over DDPG when the time constraint is more stringent. In other words, TD3 suits better under a shorter mission duration, since it has a more complex network structure and can generate more randomness to test the action space more extensively for better solutions.

Fig.~\ref{fig.cdf_dis} plots the cumulative distribution function (CDF) of the distance between the UAV and its expected destination at the end of the mission, for mission durations of $T=30$ and $40$~s. The results show that as the number of RIS elements (i.e., $N$) or the mission duration (i.e., $T$) increases, the proposed TD3 algorithm can get closer to or reach the destination more frequently, and performs significantly better than the case without an RIS. While perfect CSI is important for UAV trajectory planning, as seen in Baseline 2 for $N=20$, the TD3 algorithm can produce equally effective trajectories without CSI by utilizing a larger RIS with more elements (as seen in TD3 for $N=40$). On the other hand, the DDPG algorithm appears to suffer from overfitting, as the distance between the UAV and its expected destination has little dispersion. This indicates that DDPG is more prone to overestimating the $Q$-value function for a small number of possible actions, leading to a noisy gradient for policy refreshes and less effective UAV trajectories and lower data rates.

\section{Conclusion}
This paper developed a new DRL-driven framework for the trajectory planning and RIS-assisted jamming rejection for a UAV-borne IoT platform.
The DDPG model and its enhancement, TD3, were designed to allow the UAV to learn its trajectory and the RIS configuration
only based on its received data rate, eliminating the need of CSI for learning.
Extensive simulations showed that the proposed DRL algorithms offer the UAV reliable
resistance against jamming.
The TD3 algorithm converges faster and more smoothly than the DDPG algorithm.
It also demonstrates robustness against different locations of the jammer.
This is particularly important due to the difficulty in locating the jammer in practice.

\bibliographystyle{IEEEtran}
\bibliography{irseve_ref}
\end{document}